\newif\ifanonymous
\newif\ifdraft

\anonymousfalse
\draftfalse
\InputIfFileExists{localflags}{}

\documentclass[compsoc, conference, a4paper, 10pt, times]{IEEEtran}
\usepackage{microtype}

\ifdraft
\fi

\usepackage{versions}

\usepackage{xr}

\includeversion{conf}
\excludeversion{full}

\usepackage{amsthm}
\usepackage{amsmath}
\usepackage{amssymb}
\usepackage{stmaryrd} 
\usepackage{listings}
\usepackage{hyperref}
\usepackage{booktabs}
\usepackage{makecell}
\usepackage{xspace}
\usepackage{xcolor}
\usepackage{forest}

\usepackage[capitalize]{cleveref} 
\usepackage{mathtools}
\microtypecontext{spacing=nonfrench} 

\usepackage[
	backend=biber,
    sortlocale=en_GB,
    natbib=true,
    url=true,
    doi=false,
    eprint=false
]{biblatex}

\AtEveryBibitem{
 \clearlist{address}
 \clearfield{eprint}
 \clearfield{isbn}
 \clearfield{issn}
 \clearlist{location}
 \ifentrytype{book}{
  \clearname{editor}
  }{}
\ifentrytype{inproceedings}{
  \clearname{editor}
  \clearfield{volume}
  \clearfield{url}
 \clearfield{month}
 \clearfield{date}
}{}
\ifentrytype{incollection}{
  \clearname{editor}
  \clearfield{volume}
  \clearfield{url}
 \clearfield{month}
 \clearfield{date}
}{}
\ifentrytype{proceedings}{
 \clearname{editor}
 \clearfield{volume}
 \clearfield{url}
 \clearfield{month}
 \clearfield{date}
}{}
\ifentrytype{article}{
 \clearfield{url}
 \clearfield{month}
 \clearfield{date}
}{}
}

\newcommand{\mathcmd}[1]{{\normalfont\ensuremath{#1}}\xspace}

\newcommand{\mathname}[1]{\mathcmd{\text{\textrm{#1}}}}

\newcommand{\mathfun}[1]{\mathcmd{\mathit{#1}}}

\newcommand{\mathvalue}[1]{\mathcmd{\mathit{#1}}}

\newcommand{\mathset}[1]{\mathname{#1}}

\newcommand{\textop}[1]{\relax\ifmmode\mathop{\text{#1}}\else\text{#1}\fi}

\makeatletter
\DeclareRobustCommand{\defeq}{\mathrel{\rlap{%
  \raisebox{0.3ex}{$\m@th\cdot$}}%
  \raisebox{-0.3ex}{$\m@th\cdot$}}%
  =}
\DeclareRobustCommand{\eqdef}{=\mathrel{\rlap{%
  \raisebox{0.3ex}{$\m@th\cdot$}}%
  \raisebox{-0.3ex}{$\m@th\cdot$}}%
  }
\makeatother

\newtheorem{definition}{\textbf{Definition}}
\newtheorem{example}{Example}

\theoremstyle{remark}

\DeclareMathOperator*{\dom}{dom}

\newcommand{\ledot}{\mathrel{\ooalign{\hss\raise.200ex\hbox{$\cdot$}\hss\cr$\le$}}}
\newcommand{\gedot}{\mathrel{\ooalign{\hss\raise.200ex\hbox{$\cdot$}\hss\cr$\ge$}}}
\newcommand{\pto}{\mathrel{\rightharpoonup}} 

\newtheoremstyle{component}{}{}{}{}{\itshape}{.}{.5em}{\thmnote{#3}#1}
\theoremstyle{component}

\newcommand{\set}[1]{\{#1\}}

\newcommand{\fname}[1]{\lstinline[columns=fixed]{#1}}
\newcommand{\vname}[1]{\lstinline[columns=fixed]{#1}}
\newcommand{\tvname}[1]{\text{\lstinline[columns=fixed]{#1}}}

\definecolor{deepblue}{rgb}{0,0,0.5}
\definecolor{deepred}{rgb}{0.6,0,0}
\definecolor{deepgreen}{rgb}{0,0.5,0}

\lstdefinestyle{python}{
    language={python},
    mathescape=true,
    basicstyle=\ttfamily,
    keywordstyle=\bfseries\color{deepblue},
    emphstyle=\tfseries\color{deepred},    
    stringstyle=\color{deepgreen}
}

\newcommand{\attrwrap}{\texttt{CKA\_WRAP}}
\newcommand{\attrunwrap}{\texttt{CKA\_UNWRAP}}
\newcommand{\attrdec}{\texttt{CKA\_DECRYPT}}
\newcommand{\attrenc}{\texttt{CKA\_ENCRYPT}}

\newcommand{\attrextr}{\texttt{CKA\_EXTRACTABLE}}
\newcommand{\attrmod}{\texttt{CKA\_MODIFIABLE}}
\newcommand{\attrtrust}{\texttt{CKA\_TRUSTED}}
\newcommand{\attrwwt}{\texttt{CKA\_WRAP\_WITH\_TRUSTED}}

\newcommand{\mathpred}[1]{\mathfun{#1}}

\newcommand{\TermAnn}{\mathset{TermAnnotation}}
\newcommand{\FunAnn}{\mathset{FSA}}
\newcommand{\HornAnn}{\mathset{HCA}}
\newcommand{\iknows}{\mathpred{iknows}}
\newcommand{\storedkey}{\mathpred{storedkey}}
\newcommand{\exportable}{\mathpred{exportable}}
\newcommand{\enc}{\mathfun{enc}}

\newcommand{\Terms}{\mathcal{T}}
\newcommand{\Vars}{\mathcal{V}}
\newcommand{\Sign}{\Sigma}
\newcommand{\Strings}{\mathset{Strings}}
\newcommand{\ftrue}{\mathfun{true}}
\newcommand{\ffalse}{\mathfun{false}}

\newcommand{\freshLit}{\mathvalue{flit}}

\newcommand{\termAnn}{\tau}
\newcommand{\hornAnn}{\theta}
\newcommand{\sem}[1]{\llbracket #1\rrbracket}
\newcommand{\sfk}{\mathsf{k}}
\newcommand{\HornClauses}{\mathcal{R}}

\newcommand{\clauseDecrypt}{
\iknows(\enc(k_1,k_2)) \wedge \iknows(k_1) 
\Rightarrow 
\iknows(k_2)
}

\begin{document}

\ifanonymous 
\author{}
\else
\author{%
\IEEEauthorblockN{%
    Julian Biehl and Robert Künnemann
}
\IEEEauthorblockA{CISPA Helmholtz Center for Information Security\\
Saarland Informatics Campus}
}
\fi

\title{%
    Adaptive Exploit Generation against Security APIs
}

\maketitle

\begin{abstract}
    Proof-of-concept exploits help demonstrate software vulnerability
beyond doubt and communicate attacks to non-experts.
But exploits can be configuration-specific, for example when
in Security APIs, where keys are set up specifically for the
application and enterprise the API serves.
In this work, we show how to automatically derive proof-of-concept
exploits against Security APIs using formal methods.

We extend the popular protocol verifier ProVerif
with a language-agnostic template mechanism.
Employing program snippets attached to steps in the model,
we can transform attack traces (which ProVerif typically finds
automatically) into programs.
Our method is general, flexible and convenient.
We demonstrate its use for the W3C Web Cryptography API, for PKCS\#11
and for the YubiHSM2, providing the first formal model of the latter.

\end{abstract}

\section{Introduction}
\label{sec:intro}

Exploits are small pieces of software that take advantage of
software vulnerabilities to cause unintended behavior.
They have legitimate uses; 
proof-of-concept exploits (short: PoCs) 
show the feasibility of an attack beyond any doubt, but do
not exploit a vulnerability otherwise.
Frameworks like Metasploit~\cite{MetasploitWebpage} streamline the configuration and
execution of exploits (including the addition of payload) allowing users to select exploits from
a library.
These exploits are small scripts whose configuration is limited to a few
operational parameters,
e.g., the target host and port.
%
Such exploits can only attack software in specific configurations. If
the software is configured differently than expected, the exploit may
easily fail.

Sometimes, exploits need to adapt to the system configuration, for instance, when considering
logical attacks against Security APIs. Security APIs are
interfaces to hardware and software that provide controlled access to
security-critical data such as cryptographic keys~\cite{secapifaq}.
Prominent examples include
the  Public-Key Cryptography Standard (PKCS)\#11~\cite{pkcs11},
    used to operate hardware security modules (HSM) and
    cryptographic libraries
Microsoft's Next Generation Cryptography API~\cite{mscng}, featured on
Windows Systems since 2007,
and the
W3C Web Cryptography API~\cite{webcryptoapi}, implemented by all web
browsers to enable cryptographic operations in web applications.
These APIs have in common that they manage many keys with
individually assigned permissions. 
Due to design errors in the API,
it is sometimes possible
to extract sensitive keys simply by
combining legitimate API calls that depend on the available keys and
their permissions.
Certain attacks on PKCS\#11~\cite[Experiment~6]{DKS-jcs09}, for instance,
combine six API calls on three
keys with different permission sets. 

A single, static PoC is thus inadequate. Instead, the exploit
needs to adapt to the configuration it encounters. We can consider this task like
writing a second-order PoC: a program that, depending on the
configuration, outputs a static PoC adapted to the
situation. The challenge is to provide a method that
(a) generalizes to many different APIs,
(b) can be automated by domain experts with reasonable effort 
(c) is convenient to use
and
(d) provides flexibility in the choice of the programming language to
match the language of the API.
We propose a framework that simplifies the generation of
adaptive exploits by combining ProVerif, a popular tool for protocol
verification, with a simple template programming technique for program
synthesis. 

ProVerif is a protocol verifier with a high degree of
automation~\cite{proverif}. It operates on an abstract model of the respective
API. This task is well understood; \cref{sec:related} discusses a large body of work on 
the modeling of Security APIs and 
how to adapt protocol verifiers to
their analysis.
We annotate the attacker rules
with code-generating templates.
When ProVerif finds an attack---otherwise, there is nothing to
demonstrate---it internally represents the attack as a derivation
tree  whose edges
correspond to instances of these rules, mapping variables that occur
in these rules to concrete values.
We allow their corresponding templates to contain the same variables,
so they can likewise be instantiated.
With this operation, we can project the attack tree onto
a tree of program statements. 
Any topological sorting results in a program that respects the causal
dependencies encoded in this tree and thus a PoC.

This method is simple to use and versatile, as we demonstrate on three case
studies, 
PKCS\#11, the YubiHSM2 API~\cite{yubihsm} and the 
Web Crypto API~\cite{webcryptoapi}.
We find our method applicable to all three:
our PoCs worked in all our experiments. 
The attack finding never takes longer than five minutes and almost
always seconds. 

We list as our contributions as follows:
\begin{enumerate}
    \item An extension of the ProVerif protocol verifier that can
        produce PoC exploits from annotated model files.
    \item A mathematical model of the computed output, providing
        a basis for formal reasoning about the PoC generation.
    \item Formal models (including annotations) of the aforementioned
        APIs, including the first formal model of the YubiHSM2.
\end{enumerate}

\section{PoC exploits: what is the use?}

Exploit collections like Metasploit are employed by legitimate
penetration testers
and
cyber criminals alike.
Like any method for exploit generation, our  proposal can be adapted to
carry dangerous payload. 
The downside is obvious, so what value does the generation of PoC
exploits offer  to offset this negative
potential?

\paragraph{PoCs are communication tools.}
The biggest argument is that PoCs help non-experts understand
security-relevant issues. 
Sometimes it can be necessary to convince operators
that their systems are affected without accessing their systems. In
other circumstances, operators are convinced that an attack exists,
but need to convince management that the matter is urgent. Here, a PoC
provides irrefutable evidence that the API attack affects the system,
even to those who do not understand how the system works.
All popular vulnerability databases provides links to PoCs when
available (e.g., \url{nvd.com}, \url{vuldb.com}), some even link
Metasploit scripts (e.g., \url{cvedetails.com}).

\paragraph{PoC make analysis results accessible.} 

Our case studies analyze hardware and software configuration using
formal methods, but the generated PoCs are accessible to system operators even if they are unfamiliar
with formal methods. Interpreting the formal representation of an attack usually
requires mathematical background and experience, an entirely 
different terminology (e.g., modelers would label the `client' role
in TLS with the more general term `initiator'). The models often involve
non-trivial abstractions that need verification-specific expertise to
be correctly interpreted~\cite{nguyenAbstractionsSecurityProtocol}.

\paragraph{PoCs are trivially sound.}
Taking communication aside, such evidence can be useful to bridge the
gap between the model and the actual system. Verification tools like
ProVerif perform abstraction and may thus report attacks that are not
applicable in the model. Moreover, the model itself is almost always
an abstract representation of the system. 
An attack in the model that is unrealizable 
would lead to a failure during the PoC execution. This can be used
to improve the precision of verifier or the model. Our template
mechanism could therefore be a building block in a counter-example
guided refinement process~\cite{clarkeCounterexampleGuidedAbstractionRefinement2000}.

\section{Related Work}
\label{sec:related}

Steel~\cite{secapifaq} 
and Bond and Anderson~\cite{bond01api}
characterize Security APIs as the interface between an untrusted process
and a trusted process providing  access to sensitive resources in
a secure way. While Security APIs do not necessarily
provide cryptographic functionality, many of them (including our case
studies) do.

We can trace logical attacks on security APIs back to 
Longley and Rigby~\cite{longley92automatic}. The vulnerability they identify was later
found to affect an HSM used in the ATM network~\cite{focardiFormallyVerifiedConfiguration2021}.
PKCS\#11 in particular has received a lot of attention~\cite{clulowSecurityPKCS112003,cortier07TACAS,DKS-jcs09,bortolozzo2010,dax2019,focardiFormallyVerifiedConfiguration2021},
but  researchers have also proposed new
APIs~\cite{Kremer2011Security-for-Ke,Cortier:2012:RLL:2382196.2382293,KKS-esorics13}.
The YubiHSM API has been analyzed in conjunction with its counterpart,
the Yubikey authentication token~\cite{künnemannsteel2013,gonzalezburgueno2018}
discovering a design flaw that was subsequently fixed, as well as
showing how to set the device up securely despite this flaw. 
Both analysis build on the same model~\cite{künnemannsteel2013} but
use different tools. The YubiHSMs API was completely overhauled with
version 2, elevating it from an add-on to the token infrastructure
to a general-purpose hardware-security module. In this work, we are
the first to evaluate the YubiHSM2 which, from the modeling
perspective, is essentially a new product.
The WebCrypto API was analyzed Cairns et al.~\cite{cairns2017}, who
discovered design flaws, most of which W3C removed before
standardization.
%

\paragraph{Security API analysis}
Formal analysis of Security APIs has been conducted using
linear logic~\cite{fröschle2011},
logic programming~\cite{longley92automatic},
model checking~\cite{cairns2017,bortolozzo2010}
and 
typing~\cite{adao2013}. 
The most prominent approach employs protocol verification tools like 
ProVerif~\cite{proverif} (e.g., \cite{chevalLittleMoreConversation2018})
or
Tamarin~\cite{tamarin} (e.g.,
\cite{dax2019,focardiFormallyVerifiedConfiguration2021}).
The idea is to consider the trusted process as the only party in
a protocol where incoming messages represent API calls and the network
attacker represents the untrusted process trying to exploit this API.
Protocol verifiers are well suited to handle the
cryptographic primitives present in Security APIs and can (as opposed to model checkers) provide results for an unbounded number of keys.
Security APIs have motivated the development 
of analysis techniques and abstractions for mutable state in both 
tools~\cite{chevalLittleMoreConversation2018,
kremerAutomatedAnalysisSecurity
}. 

We picked ProVerif over Tamarin because it is (in general) faster in
finding attacks. In contrast to Tamarin, ProVerif may output false
attacks. We explore how this affects our models in
\cref{sec:applications}.
Nevertheless, our template-base attack recovery translates directly to
Tamarin's multiset-rewrite calculus. Tamarin represents
attacks as
dependency graphs with rule instantiations as nodes and causal
dependencies as edges.
Like  in the present work, the variable instantiations can be used
to instantiate the template, and a topological sorting of the graph
yields the order of statements.

Coming back to the other approaches, we see that our method
also applies to the aforementioned works building on
logic programming~\cite{longley92automatic}
and
model checking~\cite{cairns2017,bortolozzo2010}.
For model checking, depending on the concrete framework, it may be
necessary to encode the variable assignment in the API states, so they
can be easily extracted. 
Our proposal is incompatible with typing~\cite{adao2013}, because
failure to type cannot always be mapped to a counterexample.
We cannot comment on the compatibility with the 
linear logic approach~\cite{fröschle2011}, as it has not been
automated yet.

\paragraph{Attack recovery}

To our knowledge, the only work that recovers attacks on Security APIs
from formal models is by Bortolozzo et al.~\cite{bortolozzo2010}.
They use the PKCS\#11 API to poll information about the connected HSM
to create a model (reverse engineering). A model checker finds attacks (or proves
security) which  they then translate to PoCs. 
The implementation of this procedure, Tookan, was later
commercialized~\cite{StartupsResultingProject}.

Our framework generalizes this technique to arbitrary
APIs and provides a theoretical underpinning for the translation step
from the abstract attack to the concrete PoC. Our model of PKCS\#11 is
formalized in ProVerif, and hence can cover security for unbounded
number of keys. In theory, Tookan's model checking approach can only
cover bounded attacks; however, it is possible (albeit not formally
proven) that their bounded model is complete w.r.t.\ attacks.

\section{Preliminaries}
\label{sec:preliminaries}
 \label{terms}

ProVerif provides four input formats,
typed and untyped applied-pi calculus,
and typed and untyped Horn clauses.
Protocols described in the applied-pi calculus are internally
translated into Horn clauses.
Variables introduced in this translation are hidden at the applied-pi
level. We chose to extend the Horn clause format so the modeler
has full control over the variables (which may appear in the templates)
and so that the conversion is more transparent.
We picked the typed Horn clause format, because it can express untyped Horn clauses
by using the same type everywhere. 

\paragraph{Term algebra}

The smallest part of a Horn clause is a \textit{term}. 
We use terms to model messages that are sent between protocol parties, or in our case, the attacker and the Security API.
A term is either a \textit{variable}, a \textit{name},
or a \textit{function application}.
%
Variables represent other terms in Horn clauses and have a type. Names  represent atomic values like keys or identifiers. They also have a type. We denote the set of variables $\Vars$.
%
%
A function application combines a \emph{function symbol} $f$ 
with a sequence of terms and is written
$f(t_1,\ldots,t_n)$.
Each function symbol is assigned an arity $n$ defining the length of 
$(t_1,\ldots,t_n)$.
As a shortcut, we write $f/n$ for the function symbol $f$ with arity
$n$.

\begin{example}
The function symbol $\enc/2$ models encryption.
Given a term $m$ and a name $k$, $\enc(m,k)$ symbolically represents
the result of encrypting the message represented by $m$ with the key
represented by $k$.
\end{example}

A substitution $\sigma$ is a function from variables to terms. We
lift the application of substitutions from variables
to terms and use postfix notation, i.e., we write $t\sigma$ for the
term in which we replace each variable $x$ occurring in the domain of
$\sigma$ and in $t$ by $\sigma(x)$.

Names can also be created depending on the values of other term.  A ProVerif model
can only refer to a bounded number of names, but by `creating' names
as functions on, e.g., adversarial input, we can represent an
unbounded number of keys. Although such names are functionally similar
to function symbols, we denote them with square brackets. e.g.,
a name $\mathsf{handle}$ that depends on two variables $x$ and $y$ is
written
$\mathsf{handle[}x,y]$.


\paragraph{Horn clauses}
Horn clauses are built from \textit{facts}. A fact can either be a predicate over a sequence of terms, or it can be an inequality between two terms. Predicates are used to model the attacker knowledge or the content of a key storage. For example, the fact $\iknows(k)$ 
models that the intruder knows a key $k$.

Using these facts, Horn clauses  either consist of a single fact $F$, which is just a statement that this fact is true, or they can have the form $F_1 \wedge \dotsb \wedge F_n \Rightarrow F$, stating that, if all facts $F_1$ to $F_n$ are true, then $F$ is also true. 
We will typically use the letter $R$ to represent a Horn clause (a
`rule'), $H$ for the hypothesis and $C$ for the conclusion. 

\begin{example}
The following clause allows the attacker to encrypt a message: 
\[\iknows(k) \wedge  \iknows(m) \Rightarrow  \iknows(\enc(k,m))\]

The facts to the left of the arrow (called hypotheses)
state that to use encryption,
the attacker needs to know a key $k$ and a message $m$.
If both facts are true, then the fact on the right (the conclusion) is also true and the attacker knows the encryption of message $m$ with the key $k$.
\end{example}

In the typed Horn clause input, each variable and name are
additionally assigned a type, written, e.g., $k\colon \mathit{key}$.
For readability, we will omit the types unless they are relevant to
the discussion.

\paragraph{Queries}

ProVerif queries consist of the \texttt{query} keyword and a single fact that can contain variables. For instance, the following query checks if the attacker can learn a certain key.
\[\mathsf{query}\text{ }\iknows(\mathsf{key1}[])\]

\paragraph{Running example}

\cref{fig:runningexample} shows our running example, a simplified
version of the YubiHSM2 API. (We will present this API in more detail in
\cref{yubihsm}.) The YubiHSM2 can store keys, but as storage is
limited, \emph{key wrapping} is employed to securely store keys outside
the device, by encrypting them with another key. To import, that same
key decrypts the cyphertext from the previous command (which
is also called the `wrapped key'). Instead of outputting the wrapped key in the clear, it is copied into storage.
For the simplified YubiHSM2, we assume only two types of key, wrap keys and HMAC keys. These key types are represented in the model as constants, i.e., function symbols with arity 0.

The Horn clauses use three predicates.
The predicate $\iknows$ models the intruder's knowledge.
The predicate $\storedkey$ models that a certain key is stored on the device. It has arity 3 and takes the first parameter to be a handle to the key, the second to be its cryptographic value and the third to be its type (wrap key or HMAC key).
In the context of security APIs, the handle is the reference the
outside process uses to identify sensitive data they want to access
via the API (but are not allowed to learn). For the YubiHSM2, the
handle is a running number but for other tokens, it can be a memory
address or a key's hash. 
The third predicate, $\exportable$, is set if the handle
has the \texttt{exportable} attribute on the device, which governs whether it
can be exported via wrapping.

Using these predicates, we define clauses that model the interesting parts of the API. The first two (after the initial state) model the adversary's  capabilities to compute wrappings, i.e., encryptions $\enc(k,m)$, and to decrypt them (if $k$ is known).
The next two clauses model the device's functionality to export and import
keys if a wrap key is already in place. The third models the YubiHSM2's
\fname{put} command, allowing to store a key generated outside the
device (and thus known by the untrusted outside process) on the
device.

The initial configuration of the API and the initial attacker knowledge
are expressed in clauses that have no hypotheses.
In contrast to the previous rules, the initial configuration is
described using names instead of variables, 
modeling concrete keys and handles.

Finally, a query for the secrecy of the key $\mathsf{key1[]}$ is formulated, which is on the YubiHSM2, but initially unknown to the adversary.

\begin{figure}
    \newcommand{\mytitle}[1]{\textbf{\textit{#1}}}

\textbf{\textit{encrypt wrap:}} \\
$\iknows(k_1) \wedge \iknows(k_2) \Rightarrow
\iknows(\enc(k_1,k_2))$ \medskip

\textbf{\textit{decrypt wrap:}} \\
$\clauseDecrypt$ \medskip

\textbf{\textit{export wrapped:}} \\
$\storedkey(h_1,k_1,\mathsf{wrapkey)} \wedge \storedkey(h_2,k_2,t) \\
\wedge \exportable(h_2) \Rightarrow \iknows(\enc(k_1,k_2))$ \medskip

\textbf{\textit{import wrapped:}} \\
$\storedkey(h_1,k_1,\mathsf{wrapkey)} \wedge \iknows(\enc(k_1,k_2))\\
\Rightarrow \storedkey(h_2,k_2,\mathsf{wrapkey})$ \medskip

\textbf{\textit{put wrapkey:}} \\
$\iknows(k_1) \Rightarrow \storedkey(h_1,k_1,\mathsf{wrapkey})$ \medskip

\textbf{\textit{initial state:}} \\
$\storedkey\mathsf{(handle1[],key1[],hmackey)}$ \\
$\exportable(\mathsf{handle1}[])$ \\
$\mathsf{iknows(key2[])}$ \medskip

\textbf{\textit{query:}} \\
$\mathsf{query}\text{ }\mathsf{iknows(key1[])}$
  \caption{Example model.}
  \label{fig:runningexample}
\end{figure}

\paragraph{ProVerif's resolution}

When ProVerif is run with a typed Horn clause model, it will iterate
through all the queries it contains, checking for each if the
queried fact is derivable. A fact is derivable if there is
a derivation tree witnessing this fact. Valid steps in this derivation
are rules that are \emph{subsumed} by those in the model.

\begin{definition}[Subsumption~\cite{BlanchetBook09}]
We say that $H_1 \Rightarrow C_1$ subsumes $H_2 \Rightarrow C_2$, and we write $(H_1 \Rightarrow C_1) \sqsupseteq (H_2 \Rightarrow C_2)$, if and only if there exists a substitution $\sigma$ such that $\sigma C_1 = C_2$ and $\sigma H_1 \subseteq H_2$ (multiset inclusion). We  write $R_1 \sqsupseteq_\sigma R_2$ for $\sigma$ a substitution that fulfills this condition.
\end{definition}

We slightly modified \citeauthor{BlanchetBook09}'s definition to make
the substitution explicit, as we will use these substitutions to
instantiate the templates.
When presenting an attack, ProVerif provides us with a derivation tree
structured as follows:

\begin{definition}[Derivability~\cite{BlanchetBook09}]\label{def:derivability}
Let $F$ be a closed fact, that is, a fact without variable. Let $\HornClauses$ be a set of clauses. $F$ is derivable from $\HornClauses$ if and only if there exists a derivation of $F$ from $\HornClauses$, that is, a finite tree defined as follows:
\begin{enumerate}
\item Its nodes (except the root) are pairs of clauses $R \in \mathcal{R}$ and substitutions $\sigma$;
\item Its edges are labeled by closed facts;
\item If the tree contains a node $(R, \sigma)$ with one incoming edge labeled by $F_0$ and $n$ outgoing edges labeled by $F_1, . . . , F_n$, then $R \sqsupseteq_\sigma F_1 \wedge. . .\wedge F_n \Rightarrow F_0$.
\item The root has one outgoing edge, labeled by $F$. The unique son of the root is named the \textit{subroot}.
\end{enumerate}
\end{definition}

Here we modified \citeauthor{BlanchetBook09}'s definition (and adapted
ProVerif) to include
the substitution $\sigma$ in the node's annotation (instead of the
subsumed rule).  

\begin{example}
    \newcommand{\ka}{\mathsf{key1}[]}
    \newcommand{\kb}{\mathsf{key2}[]}

    The following partial derivation tree 
    for \cref{fig:runningexample} outlines how 
    $\iknows(\ka)$ can be derived 
    by decrypting a wrapping under key $\kb$. (The part where $\kb$ is  
    imported is omitted.)
    
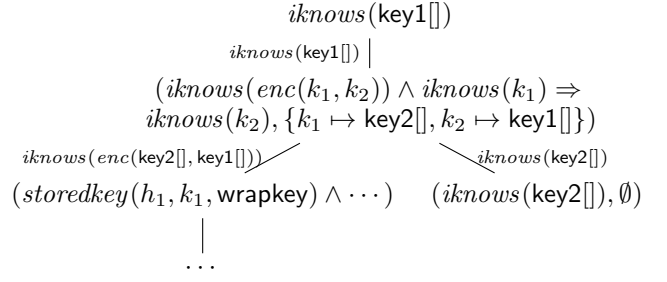
\begin{center}
    \begin{forest}
       [$\iknows(\ka)$
       [
       \parbox{7cm}{\centering
       $(\clauseDecrypt,\{k_1 \mapsto \kb, k_2 \mapsto \ka \})$},
       edge label={node[midway,left,font=\scriptsize]{$\iknows(\ka)$}}
           [{$(\storedkey(h_1,k_1,\mathsf{wrapkey)} \wedge \cdots)$}, edge label={node[midway,left,font=\scriptsize]{$\iknows(\enc(\kb,\ka))$}}
                [{$\cdots$}]
           ]
           [{$(\iknows(\kb),\emptyset)$}, edge label={node[midway,right,font=\scriptsize]{$\iknows(\kb)$}}]
       ]
       ]
    \end{forest}
    \end{center}
\end{example}

\section{Approach}
\label{sec:approach}

To translate an attack trace created by ProVerif into a PoC, we map
each Horn clause application in it to code that
realizes it.
We extend the  typed Horn clause modelling language with annotations that carry the necessary information.
We implemented this feature as an extension to ProVerif. Besides improving the
user experience, this allows us to rely on a lot of features and
internal data structures that ProVerif offers.

In Section \ref{translation}, we define how we combine the input model's annotation
to produce PoCs from attack traces.
Then we describe the template format, give some implementation details
and formulate some modeling guidelines that we adhered to in our case
studies.

\subsection{Translation} \label{translation}

Let $\Terms$ be the set of terms we defined in \cref{terms}. 
We annotate Horn clauses with templates of the following form.

\begin{definition}
A \emph{Horn clause annotation} is a string from an alphabet that
includes a set of distinguished symbols 
$\cdot_t$ for each $t\in\Terms$. 
We call the set of Horn clause annotations
$\HornAnn$. The set of strings that do not contain any distinguished
symbols is called $\Strings \subseteq \HornAnn$.
\end{definition}

We do not model how these symbols are encoded, but assume a function
that can replace terms with strings.

\begin{definition}[Horn clause substitution]
For a partial function $\alpha \colon \Terms \rightharpoonup \HornAnn$,
and a Horn clause annotation $a$, let 
$\sem{a}^\alpha \in \Strings $  
denote $a$ where each $\cdot_t$ such that $t\in\dom(\alpha)$
is substituted by $\alpha(a)$.
\end{definition}

The idea is that each Horn clause that represents an attacker
action carries such an annotation.
Some Horn clauses represent internal actions of the trusted process; these are not annotated.
We nevertheless need them to find attacks, as they reflect the behavior of the system,
but no code in the PoC represents them.
Other Horn clauses model the interaction of the trusted process with
the attacker or computations performed by the attacker alone.
They are both important to find attacks and are represented in the
PoC.

\begin{example}
This Horn clause models the attacker's capability to compute symmetric encryption:
\[\iknows(k)\land \iknows(m) \Rightarrow \iknows(\enc(k,m))\]
To produce python code that realizes this action, i.e., produces this ciphertext,
we use the following annotation $a$: 
\begin{lstlisting}[style=python]
    $\cdot_{\enc(k,m)}$ = $\cdot_k$.encrypt($\cdot_m$)
\end{lstlisting}
The substitution typically maps the terms to the variable identifiers
that store the result of this computation. For example, the mapping
\[
    \alpha = \{ 
        k \mapsto \tvname{key1}, \quad
        m \mapsto \tvname{msg}, \quad
    \enc(k,m) \mapsto \tvname{c} \}
\]
    
would produce the string `\lstinline[style=python]{c = key1.encrypt(s1)}'.
This example illustrates why $\alpha$'s domain is the set of terms
instead of the set of variables: the PoC needs to refer to intermedeate computation results for readability.
\end{example}

To obtain the substitution in a systematic way, we also
 have to translate terms into strings.
To this end, we annotate each function symbol with a template.
Function symbol annotations work similarly to Horn clause annotations,
but as function symbols have no variables, they identify positions in
the term via integer symbols.

\begin{definition}[Function symbol annotation]
    We annotate a
function symbol with arity
$n$
with a string from an alphabet that
includes a set of distinguished symbols 
$\cdot_n$ for each $n\in \{1,\ldots,n\}$.
We call the set of function symbol annotations
$\FunAnn$.
\end{definition}

\begin{example}\label{ex:term-annot-attr}
    In our PKCS\#11 model, we encode the attributes 
    with a function symbol $\mathsf{attributes}/4$
    that is annotated with the following simple template. 
\begin{lstlisting}[style=python]
    processAttributes($\cdot_1$,$\cdot_2$,$\cdot_3$,$\cdot_4$)
\end{lstlisting}
    This function constructs a list of attributes with the proper type.
\end{example}

Term annotations are used to translate terms into what they represent
in the PoC, e.g., key handles, cryptographic objects, attribute
objects.
Terms are translated recursively; hence we also need to define how to
translate names and variables.
In both cases, we chose a fresh literal.\footnote{This comes at a slight
    loss of generality: the template language assumes the programming
language to have literals described as ASCII letters.}
For names, this is the natural representation.
For variables, this is equivalent to instantiating the variable with a fresh
name.
%
%
Our implementation preserves a counter to ensure each literal is unique. To abstract away from our naming scheme and its
statefulness, we instead
assume some injective function $\freshLit : \Terms \to \Strings$
that choses a fresh string for each term.

\begin{definition}[Term translation]
Let
$\termAnn \colon \Sign \pto \TermAnn$ 
be a partial function 
from function symbols to term annotations,
and
$t$ a term.
Then, 
    $\sem{t}^\termAnn$ is defined as follows.
    If $t$ has form $f(t_1,\ldots,t_n)$
    and $\termAnn(f)$ is defined,
    it is  $\termAnn(f)$ with 
    with $\cdot_i$ replaced by
    $\sem{t_i}^\termAnn$, for all $i\in\set{1,\ldots,n}$.
    Otherwise, it is $\freshLit(t)$.
\end{definition}

Note that $\Sign$ includes only function symbols that we explicitly
declare, e.g., $\mathsf{key}[\mathsf{name}[]]$ is a name
and therefore assigned a fresh literal.

\begin{example}
    Picking up from 
    \cref{ex:term-annot-attr},
    we define
    $\termAnn$ to
    assigning to the function symbol 
    $\mathsf{attributes}/4$
    the aforementioned template,
    and to the function symbols 
    $\ftrue/0$
    and
    $\ffalse/0$ the string annotations 
    \vname{True}
    and
    \vname{False}.
        Then, the term 
    $\mathsf{attributes}(\ftrue,\ftrue,\ffalse,\ftrue)$ translates to
\begin{lstlisting}[style=python]
    processAttributes(True, True, False, True)
\end{lstlisting}
\end{example}

If ProVerif finds an attack, we obtain a derivation tree proving that
the queried term is indeed derivable. Recalling \cref{def:derivability}, the nodes
in this tree have labels $(R, \sigma)$, where $R$ is a Horn clause 
possibly associated to a template and 
$\sigma \colon \Vars \pto \Terms$ an instantiation of the variables 
in the Horn clause. Thus $\sigma$ is also applicable to the template.
Using this translation mechanism for terms,
we are now ready to translate the nodes of this derivation tree
into a string.

\begin{definition}[Derivation step translation]
Let $\termAnn : \Sign \pto \TermAnn$
and $\hornAnn: \HornClauses \pto \HornAnn$
be the term and Horn clause annotations for a set of clauses
$\HornClauses$. 
We translate a node $(R,\sigma)$ as follows:
\[
    \sem{R,\sigma}^{\termAnn,\hornAnn} \defeq
    \begin{cases*}
        \sem{\hornAnn(R)}^\alpha
        & \text{if $\hornAnn(R)\in\HornAnn$ defined}
        \\
        \epsilon & \text{otherwise}
    \end{cases*}
\]
where $\alpha(t) = \sem{t \sigma}^\termAnn$. 
\end{definition}

In other words, we translate rules without annotations to the empty
string $\epsilon$. Rules with annotations, we translate by
composing the substitution $\sigma$ and 
the term translation w.r.t. $\termAnn$ 
and applying the result ($\alpha: \Terms \to \Strings $) to the annotation.
The following example will demonstrate how this plays out in practice:

\begin{example}
The following Horn clause $R$ represents an API call for 
importing a key $k_1$ with a set of attributes $\mathit{attrs}$:
\[\iknows\mathit(k_1) \wedge \iknows\mathit{(attrs)} \Rightarrow 
\storedkey(\sfk(k_1,\mathit{attrs}))\]

The function symbol $\sfk[]$ represents the key as stored within the
API, combining its cryptographic value and associated metadata, i.e.,
the set of attributes. It has no
term annotation, but we annotated the clause with $\hornAnn(R)=$
\begin{lstlisting}[style=python]
    $\cdot_{\sfk(k_1,\mathit{attrs})}$ = API.import($\cdot_{k_1}$, $\cdot_{\mathit{attrs}}$)
\end{lstlisting}
Now assume we want to translate a node  $(R, \sigma)$ with 
$\sigma = \{
k_1\mapsto\mathsf{key1[]},
\mathit{attr}\mapsto\mathsf{attributes(true,false,false,true)}
\}$. 
The effect of the term translation to
$\sem{\mathit{attrs} \sigma}^\termAnn = 
\sem{\sigma(\mathit{attrs})}^\termAnn$, as per the previous example,
is
\begin{lstlisting}[style=python]
    processAttributes(True,True,False,True)
\end{lstlisting}
Because
$\sem{k_1 \sigma}^\termAnn = 
\sem{\mathsf{key1[]}}^\termAnn$
and
$\sem{ \sfk(k_1,\mathit{attrs})\sigma}^\termAnn = 
\sem{ \sfk(\mathsf{key1[]},\mathsf{attributes}(\cdots))}^\termAnn$,
the term translation assigns a fresh literal each, say 
$\mathsf{x1}$
and 
$\mathsf{x2}$.
We hence translate the node to the string:
\begin{lstlisting}[style=python]
x2 = API.import(x1,processAttributes(
            True,True,False,True))
\end{lstlisting}
\end{example}

The order in which nodes are translated should respect the order of
the derivation tree. 
\begin{definition}
Let $\termAnn : \Sign \pto \TermAnn$
and $\hornAnn: \HornClauses \pto \HornAnn$
be the term and Horn clause annotations 
for a set of clauses $\HornClauses$.
For any derivation $T$ of some closed fact from $\HornClauses$,
we define the translation 
$\sem{T}^{\termAnn, \hornAnn} = 
\sem{n_1}^{\termAnn, \hornAnn} || \cdots ||
\sem{n_m}^{\termAnn, \hornAnn}$ 
for $n_1$ to $n_m$ the nodes of $T$ in some arbitrary but fixed
topological order and $||$ the concatenation operation.
\end{definition}

With this method, each node in the derivation tree is translated into
the corresponding output line
and the tree is linearized so that the lines form a program.
We implemented the linearization of $T$ in depth-first order.

\subsection{Implementation}\label{sec:implementation}

We implemented our approach in ProVerif and made its source code available as free software (see the supplemental material).
%
By extending ProVerif, instead of processing its output, we could rely
on its parser and internal data structures for more reliable processing.
Our fork offers all the functionality of the original ProVerif with an additional input mode for annotated models
and
an additional output mode for PoCs.

To use the new input mode, ProVerif is called with \texttt{-in exthorntype} or a model file that ends with \texttt{.exthorntype}. 
The new output mode is called with \texttt{-out poc} and an output path \texttt{-o $\mathit{path}$} and requires the new input mode.

\paragraph{Template format.}
Annotations are always enclosed between brackets
\texttt{(**}
and
\texttt{**)}. The symbol $\cdot_t$ or $\cdot_n$ is recognised with a special
delimiter defined as the first symbol after the opening bracket, e.g., 
\begin{lstlisting}[style=python]
(**| |k(k1,attr)| = API.import(|k1|,|attrs|) **)
\end{lstlisting}
uses \vname{|} as the term limiter.
In addition to Horn clause annotations and function symbol
annotations,
a header and footer template can be specified at the top and bottom of the
input file. They are placed at the start and end of the PoC,
respectively. They can be used to define functions, import libraries
or place code that cleans up after the attack has been executed.

\section{Applications}
\label{sec:applications}

We tested our approach with three highly popular security APIs, 
the Web Cryptography API~\cite{webcryptoapi}, 
PKCS\#11~\cite{pkcs11}
and the YubiHSM2 API~\cite{yubihsm}. For each for them, we will first present its background, then our model and finally evaluate these models to argue  
for the effectiveness of our method. 
Afterwards,  we summarize our results.

\subsection{Web Cryptography API}\label{sec:webcrypto}

The Web Cryptography API (short: `WebCrypto API') offers cryptographic operations for web applications. 
The API is provided via JavaScript and implemented in every major web browser~\cite{webcryptosupport}.
The WebCrypto API does not have stated security goals
and is flawed in many aspects.
It nevertheless provides cryptographic functionality and an indirect
access to keys; hence we consider it a suitable object to test our PoC
generator.

It offers private and public key cryptography, but also key wrapping 
and key unwrapping, the opposite operation. When unwrapping a wrapped
key, i.e., a payload key encrypted with another key, the user requests
the API to decrypt the wrapped key and store the result of the
decryption as a new key.
APIs that provide wrapping and unwrapping are susceptible for many
attacks~\cite{clulowSecurityPKCS112003}, the simplest one being
wrap-and-decrypt. If the wrapping key (=the encryption key) can also
be used for decrypting arbitrary payload, then the attacker can obtain
a key in the clear (possibly the wrap key itself!) by wrapping it and
decrypting it. 
The support for wrap and unwrap is reason enough to investigate the
WebCrypto API and indeed, \citeauthor{cairns2017} have done so in the
path. 

In WebCrypto, every key has an attribute that specifies its permitted use,
which makes it difficult to find attacks by hand.
One of them,  
`extractable'
states that the API permits extracting the key. However, key wrapping,
which can be set up securely~\cite{bortolozzo2010,dax2019,focardiFormallyVerifiedConfiguration2021}
is not the only way of extracting a key. The function \fname{exportKey} exports any key marked as extractable in the clear. We thus follow \citeauthor{cairns2017}'s approach and only consider attacks that involve key wrapping.

\subsubsection{Modeling}

The two most important parts of our model
are the predicates $\storedkey$ and $\iknows$, which have the same
semantics as in the running examples.
Using those, we modeled from the API calls described in the official
specification~\cite{webcryptoapi}:
\begin{itemize}
    \item all those related to key management, except for
        \fname{exportkey} (to avoid trivial
        attacks~\cite{cairns2017}), and
    \item encryption and decryption as key usage functions.
        For the similarly structured PKCS\#11 API, it is usual to only
        model these key usage functions as they can potentially
        interact with wrapping and unwrapping. (The argument from, e.g., 
        \cite{künnemann2015} applies directly.)
\end{itemize}

The WebCrypto API offers 
multiple key formats for wrapping that store different key metadata.
Some contain just the key, others also include the key type, the key usages or both.
The model reflects all these differences and thus includes four rules
each for wrapping and unwrapping.

While constructing the model, we noticed that the Web Crypto implementations are not behaving the same in the browsers we tested (Mozilla Firefox and Google Chrome). 
We found some aspects in which an earlier version of Firefox does not
adhere to the standard (which is documented in the supplementary
material) but were independently found and corrected.
%
%
Moreover, Firefox refuses to unwrap AES-KW keys, 
possibly for security.
%
%
There are possibly other bugs that our test cases did not
find---hunting for these was outside this work's focus, but we
suspect there are more.

The Web Crypto API itself does not contain any storage mechanism for
keys.  Instead, key objects are expected to be
stored using the Indexed
Database API so they can be used `without ever exposing that key
material to the application or the JavaScript environment'~\cite[Section~5.2]{webcryptoapi}.
As we did not experiment with any particular application, we skipped
this extra indirection and used the web crypto API directly.
Adapting to a concrete application should be straightforward, but
depends on the indexing scheme: the specification prescribes that
`the key is some string identifier meaningful to the
application'~\cite[Section~5.2]{webcryptoapi}.

\subsubsection{Evaluation} \label{webcryptoeval}

    \begin{table*}
        \centering
        \caption{WebCrypto API: Model size by number of Horn
            clauses and lines of code (LoC), PoC size  and
        recovery time.}
        \label{tab:webcrypto}
\begin{tabular}{@{}cccccccc@{}} \toprule
     & \multicolumn{3}{c}{Model Size} & & \multicolumn{2}{c}{PoC Size}\\ \cmidrule(r){2-4} \cmidrule(lr){6-7}
\makecell{Expe- \\ riment} & \makecell{\# Horn \\ Clauses} & \makecell{LoC Templates  \\ (incl. Header/Footer)} & \makecell{LoC initial\\ state} & \makecell{Stored \\ keys} & \makecell{LoC} & \makecell{\ldots excl. \\ Header/Footer} & \makecell{Attack gene-\\ ration time}
\\ \midrule
1  & 173  & 227  & 4  & 1  & 65  & 5  & 1m45s \\
2  & 173  & 226  & 3  & 2  & 66  & 6  & 1m48s \\
3  & 174  & 227  & 4  & 3  & 66  & 6  & 1m33s \\
4  & 176  & 230  & 7  & 5  & 65  & 5  & 2m15s \\
5  & 179  & 234  & 11 & 8  & 65  & 5  & 3m13s \\
6  & 183  & 240  & 17 & 12 & 65  & 5  & 5m10s \\ \bottomrule
\end{tabular}
    \end{table*}

We used our model to analyze six configurations of the API. Each was designed so that at least one key can be extracted, so that we would always find an attack and see how our approach performs. To be able to compare the results, we only looked for exactly one attack in each experiment. The results of these experiments can be found in \cref{tab:webcrypto}.

We quickly noticed that whether a key can be extracted or not only depends on whether it carries the 
\vname{extractable} attribute.
If the attribute is not set, the key is secure but entirely confined
to the device, not even wrapping is possible.
If it is set, it can be attacked.
The API provides no mechanism to avoid such an attack.
We found two kinds of attacks: in the first, the configuration 
contains a key that has both the \vname{wrapKey} and \vname{decrypt} usages.
This key can be used to wrap any other extractable key, decrypt the ciphertext and recover the wrapped key.
If no such key is present, the second kind of attack first creates
such a key, using \fname{generate_key} or \fname{import}, then
mounts the first kind of attack.

We thus designed the experiments to scale with the number of keys: for each experiment, 
we defined an initial configuration with the target number of keys with random types and usages.
We marked at least one of them as extractable, so it would be attacked. 

We consider the complexity of the model 
by the number of Horn clauses,
indicating the complexity of the modeling task,
and the 
lines of code (LoC) used for the templates,
indicating the effort for generating PoCs.
The latter number 
includes the header and
the footer, which had 60 LoC for all experiments.
The description of the initial state includes both the initial Horn
clause and a template to set up the corresponding key. It obviously
scales in the number of stored keys.
As we did not analyze an existing web application, we produced these
by hand. An actual deployment would, similar to
\citeauthor{bortolozzo2010}'s Tookan tool, generate predicates
describing an initial configuration with a script that is simply run
before ProVerif is called.

We observe that the size of the PoC does not scale with the number of
keys, but the time to generate the attack does. Still, the computation
time is reasonable.

The model carries Many more lines of code are used to annotate the model than to
describe the PoC. Moreover, the vast majority of the PoC is static, but this is mostly caused by the models header and footer annotation.
Indeed, the generated parts of the PoC are different, despite the uniformity of the attacks,
as they account for different key types and positions. A handwritten PoC would thus need to adapt the attack pattern not
only to the key handles on the device, but also their types
and usages.

In summary, we find that our methodology can be adapted to the
WebCrypto API setting, providing the first mechanized attack finder
for this scenario. Attacks can be found in about five minutes for up to 12
keys. The PoCs are not particularly complex, but the non-static part
of the code takes a different shape depending on the types and usages
of keys under attack, despite encoding only one of two attack
patterns.

\subsection{PKCS\#11}

The PKCS\#11 standard was created to provide an API to interface with security tokens independent
of whether they are implemented in software, smart cards or expansive shielded hardware modules.

The API allows for various cryptographic operations like encryption, decryption, hashing, etc. Moreover, it also supports key wrapping, which makes it interesting for our use case. Keys can be public key objects, private key objects or secret key objects. Other kinds of objects like data objects are irrelevant for attack finding. 
Each key object has a set of attributes.
Some of these are specific to key types (e.g., the public exponent is
specific to private key objects of RSA key type), others are
common to all key objects. Among the latter is
\attrextr, stating that a key can be extracted by wrapping or in the clear,
\attrmod, stating that a key can be modified
and several others specifying whether a key can be used for encryption, decryption, wrapping, unwrapping, etc.
An interesting point (sometimes considered a flaw~\cite{dax2019})
about these attributes is that they are not part of the key wrapping, so a key that is unwrapped may obtain a new and possibly conflicting set of attribute values.

PKCS\#11 as a whole is thus well known not to protect the
cryptographic value of sensitive
keys~\cite{bortolozzo2010,cortier07TACAS,DKS-jcs09,focardiFormallyVerifiedConfiguration2021},
but tokens are free to only support a subset of the operations defined by the standard.
Much of the existing work is thus about exploring secure
configurations, both in terms of the subset of operations a token
supports, or how to set the attributes of a set of keys.

\subsubsection{Modeling}

Our model closely follows that of \citeauthor{künnemann2015}. 
Again, the two most important predicates in this model are \storedkey and \iknows.
We model the API calls for encryption, decryption, key generation, key
wrapping and key unwrapping.
We distinguish public keys, private keys and secret keys. 
We associate each key with a handle so the API can tell apart
two keys that have identical cryptographic values---this situation can
occur when a key is reimported, e.g., via wrapping and unwrapping.
We further associate with each key 
a value,
a type (e.g., AES or RSA) and a set of attributes. 

First published in 1995,
the PKCS\#11 specification spans several documents with well over 400
pages, so we had to pick the functionality we wanted to support.
Just like in \cite{künnemann2015}, we only model attributes that are
relevant for key management excluding asymmetric wrapping. Asymmetric
wrapping is vulnerable to a so-called Trojan-key attack and thus
rarely supported on (well designed) tokens.
In contrast to them, we do not model wrap templates,
because we found them difficult to model in ProVerif.

Because our mechanism is language agnostic, we could use the
\texttt{python-pkcs11}~\cite{pythonpkcs11} wrapper to PKCS\#11,
providing a more convenient interface than the C API.
For the evaluation, we used the 
openCryptoKi~\cite{opencryptoki} software token.
As this token did not support the attribute
\attrwwt and \attrtrust, we had to ignore them.
%
%
Another limitation of the software token is that the only key types that can be used for encryption, decryption, wrapping and unwrapping are AES keys and DES3 keys.  We reflected this in the model. We decided that it is sufficient to model key generation for AES keys, as those can be used for all the operations mentioned above.  Generating any other type of key does not give us any advantage in finding attacks.

For convenience, we use the header to define
a function \fname{build\_template} that helps translate the attributes into a suitable object of PKCS\#11
and an array  \fname{used\_keys} that stores keys that the PoC creates
during execution.
After the operation, the footer code iterates through this array to
clear all these keys from the token. 

\subsubsection{Evaluation}

\begin{table*}
    \centering
    \caption{PKCS\#11: 
        Model size by number of Horn
            clauses and lines of code (LoC), PoC size  and
        recovery time.}
    \label{tab:pkcs}
\begin{tabular}{@{}cccccccc@{}} \toprule
    \\ \midrule
1  & 19  & 138  & 6  & 2  & 70  & 22  & 0m0,025s \\
2  & 19  & 138  & 6  & 2  & 71  & 23  & 0m0,022s \\
3  & 21  & 142  & 10 & 4  & 71  & 23  & 0m0,031s \\
4  & 23  & 146  & 14 & 6  & 70  & 22  & 0m0,033s \\
5  & 29  & 158  & 26 & 12 & 71  & 23  & 0m0,037s \\ \bottomrule
\end{tabular}
\end{table*}

Using our model, we analyzed five configurations of the API. 
As for the Web Crypto API, we designed each experiment so it has exactly one vulnerable key. The results of the experiments can be found in \cref{tab:pkcs}.

The first two are configurations used in experiments from earlier
work~\cite{delaune2008}.
%
%
In both, there are two symmetric keys stored on the device, $k_1$ and $k_2$, and the intruder knows a key $k_3$ that can potentially be used in an attack but is not stored (consequently it is not counted in \cref{tab:pkcs}).

In experiment 1, $k_2$ has the attributes \attrdec and \attrwrap which leads to an attack where $k_2$ is used to wrap $k_1$ and decrypt the ciphertext to recover $k_1$ in clear. In experiment 2, $k_2$ does not have these attributes, but instead has the attributes \attrenc and \attrunwrap. This resulted in an attack where $k_2$ was not used at all, but instead a new key was generated that could be used for decryption and wrapping, which was then used to carry out the same attack as in experiment 1. For experiments 3-5, we scaled the number of keys.


Similar to \cref{webcryptoeval},
we find the PoCs to have approximately the same size, but to be different
on the lines that are attack-specific, i.e., those outside the header
and footer.
The attack generation is extremely fast, which is faster than in the
WebCrypto API model. The annotations are also much more concise, which
is first, because JavaScript APIs tend to be verbose and second,
because the python wrapper around PKCS\#11 further simplifies the PoC.

\subsection{YubiHSM2} \label{yubihsm}

The YubiHSM is the server-side counterpart to the popular Yubikey
authentication token. Its initial purpose was to provide increased
security against server compromise while providing functionality
adapted to Yubikey's authentication protocol. 
The revised version 2 (the version we modeled and experimented with),
however, has a much wider scope, including securing cryptocurrency
exchanges, IoT environments, cloud services and Microsoft Active
Directory~\cite{yubicoabYubiHSMYubiHSMFIPS}.
It now supports 
public key cryptography, symmetric cryptography, hashing, and many
other features,
accessed via a device-specific API~\cite{yubihsmdocu}.

Each key has an ID and a type; both values combined distinguish keys.
To be able to use the API, a user has to log in with the ID of an \emph{authentication key}
and the value of that key (or a password that derives the key). 
This key is then used to establish a session. 
Every key stored on the device has a set of \textit{capabilities}, which define what the user is allowed to do with that key. For an authentication key, these capabilities define what a user can do in a session that was established with that key.

Since version 2, the YubiHSM supports key wrapping with keys of type \textit{wrap key}.
Authentication keys and wrap keys both have \textit{delegated capabilities}.
For authentication keys, these define what capabilities keys created by those authentication keys (i.e., in a derived session) can have.
For wrap keys, they define a set of capabilities that keys are allowed to have so they can be wrapped by these wrap keys.

Domains constitute another layer of access control. A key has one or more
domains. Each session is tied to a set of domains that is inherited from
the authentication key that created it.
Within such a session, the user can only use keys that share at least
one domain with the session. Furthermore, keys also need to share at
least one domain when they interact (e.g., during wrapping or unwrapping).

In summary, keys have 
ids,
types,
capabilities,
delegated capabilities
and domains.
Whether an operation succeeds may depend
on how these attributes are set (a) for the current sessions and (b) the one
or two keys involved in the current operation. 
The complexity of this state space makes is difficult to find
attacks by inspection and very attractive to validate configurations
automatically for security.

\subsubsection{The Model}

Once again, the model mostly builds on the predicates 
\storedkey
and \iknows.

For wrap and unwrap,
we model the API functions
\fname{export_wrapped} and
\fname{import_wrapped}, as well as 
\fname{generate_wrap_key} for wrap key generation
and
\fname{put_wrap_key} for storing a wrap key known to the user/attacker on
the device.

Additional clauses model the attacker's capability to craft key wraps
or to decrypt them. The YubiHSM2's own encryption and decryption
functions cannot be misused to perform these operations
---here, the YubiHSM2's designers learned from the design flaws of
PKCS\#11 and the WebCrypto API and use authenticated fields to
distinguish a wrapping from the encryption of other payload.
Our experiments confirmed that the operations can be performed outside
the device, but only if the wrap key is known to the attacker (see
\cref{sec:yubihsm-reverseengineering}).

A wrap of a key does not only contain the key itself, but also
contains all its attributes, which we refer to as the \emph{wrap
format}. This additional data is authenticated via an AEAD
scheme~\cite{rogawayProvableSecurityTreatmentKeyWrap2006}.
As we found no official documentation on the content and structure of
this data we 
reverse-engineered 
this structure using fuzzing. We could construct and deconstruct wrap formats
and output deducible keys in the clear. 
This also included HMAC keys, which are likewise not stored in plain,
but in a specific, undocumented format.
See \cref{sec:yubihsm-reverseengineering} for notes on reverse
engineering.

We model keys with a function symbol 
$\mathsf{k}/5$
combining its id, cryptographic value, type, capabilities and
delegated capabilities (which can be empty for key types that do not
support them).
The model ignores the key's size, instead the attack generation
templates transparently take them into account.

Again, we focused only on the wrapping-related functionality and the
capabilities important for them. 
We skipped the domain feature, because it did not fundamentally add
anything to the modeling approach.
Adding this aspect to the model would make it larger, but not
alter the approach at all: the domain set would provide an
additional parameter to $\mathsf{k}/7$ that matches with the session's
domain (just like its delegated capabilities).

We wrote the template code in Python~3 to interact with Yubico's SDK~\cite{pythonyubi}. 
The model relies on an authentication key provided in the initial
state via a predicate $\mathpred{iauth}$ and header code that creates
the session.\footnote{%
The attacker can potentially increase its capabilities by 
creating new
sessions from new authentication keys  (using
\fname{put_authentication_key} to put them in the store).
But the capability set of a freshly placed authentication key must be
a subset of the current session capabilities.
We hence disregard this attack vector.}
Moreover, the header contains a function that translates the
capabilities from the model into code and a stateful function that
ensures that attacker-generated wrap keys obtain unique IDs.
%
%
Again, the footer code performs cleanup, deleting keys generated by
the PoC.

\subsubsection{Evaluation}

\begin{table*}
    \centering
    \caption{YubiHSM:
        Model size by number of Horn
            clauses and lines of code (LoC), PoC size  and
        recovery time.}
    \label{tab:yubihsm}
\begin{tabular}{@{}cccccccl@{}} \toprule
    \\ \midrule
1  & 18  & 174  & 8  & 3  & 106  & 51  & 0m01s \\
2  & 18  & 174  & 8  & 3  & 146  & 91  & 0m01s \\
3  & 18  & 174  & 8  & 3  & 106  & 51  & 0m01s \\
4  & 18  & 174  & 8  & 3  & 146  & 91  & 0m01s \\
5  & 17  & 172  & 6  & 2  & 119  & 64  & 0m01s \\
6  & 17  & 172  & 6  & 2  & 119  & 64  & 0m40s \\
7  & 27  & 192  & 26 & 12 & 119  & 64  & 0m08s \\ \bottomrule
\end{tabular}
\end{table*}

We used the model to analyze the YubiHSM2 API in seven
configurations. Once again, we designed each experiment such that one
key would be vulnerable. The results are displayed in
\cref{tab:yubihsm}.

The YubiHSM2's documentation advises the user to generate an authentication key with a limited
capability set when setting up their application.
The first five experiments constitute realistic configurations where
the attacker's session is limited through the capability set of the
authentication key used to establish it.
Experiments 6 and 7 explore the impact of enlarging the capability set
and the set of keys.

The first five experiments have three keys stored on the device:
$k_1$ is an authentication key that establishes the session,
$k_2$ is a key (of arbitrary type) that has the capability \vname{exportable-under-wrap}
and
$k_3$ is a wrap key known to the attacker.
The authentication key $k_1$ (and thus the attacker's session) has all capabilities except the following:
\vname{put-wrap-key} is removed, so the attacker cannot import
a known key to produce wrappings.
Moreover, we remove the capabilities 
\vname{exportable_under_wrap} 
and
\vname{generate_wrap_key},
which forbids wrapping the
authentication key (which is useless for the attacker and most
applications) 
or creating custom wrap keys.
All other capabilities are enabled.
In experiment 2, $k_3$ has the \vname{export-wrapped} capability and the
delegated capability \vname{exportable-under-wrap}, so the attacker uses
it to wrap $k_2$ uses its knowledge of $k_3$ to decrypt the
ciphertext.
In experiment 3, $k_3$ cannot export, but instead has the capability \vname{import-wrapped}
and the necessary delegated capabilities so that a wrapping imported
with $k_3$ results in a key $k_4$ that can mount a wrap-then-decrypt
attack as in experiment 2. Because the attacker knows $k_3$, they can
create such a wrapping outside the YubiHSM2 and thus elevate their
privileges.
In both cases, attack generation takes less than a second, which is
due to the reduced attack space in this scenario.

Experiments 4 and 5 are identical to 2 and 3, respectively, except
that session established with $k_1$ can create wrap keys again (i.e.,
$k_1$ has \vname{put-wrap-key})
but has no delegated capability except 
\vname{exportable-under-wrap}.
Hence $k_1$-sessions can create wrap keys, but
none with the necessary capabilities to wrap $k_2$.
Not directly, at least:
as $k_3$ has all these delegated capabilities (and is once again known to the intruder)
the attacker can create wrap keys outside the device and import them.
We again obtain an attack in less than one second.

In experiment 6 we explore the impact of $k_1$'s capability set. We
start from experiment 1, but
give $k_1$ full capabilities. To force the attacker to make use of 
\vname{put-wrap-key}, we also remove $k_3$ from the initial store, but
still assume $k_3$ to be known.
The intruder can import $k_3$ as a wrap key, use it to wrap $k_2$ and
decrypt the ciphertext externally using $k_3$ to recover the key. 
We see that attack generation takes forty seconds, highlighting the
impact of the session capabilities on 
the attack space ProVerif has to consider. 

For experiment 7, we again start from experiment 1, but instead of
increasing the capabilities of $k_1$, we add
ten additional keys of random types.
Unsurprisingly, we find the same attack as in experiment 1, but this
time,
the resolution had to determine which of these keys were
useful.
We find that attack finding takes only marginally longer (8s),
indicating that the size of the configuration influences the
time to find an attack, but less than the capabilities of
the authentication key.

We observe that throughout these experiments,
attack generation is very fast, although the PoCs are larger than in the other examples.
The PoCs already require more than a hundred lines of code, making
automatic generation particularly attractive on this example.

\section{Discussion}

We evaluated our methodology on three security APIs and determined that, without major difficulties, 
we could write annotated models that produced valid PoCs in 
all experiments for all the APIs.
We find clear benefits for our approach for PKCS\#11 and YubiHSM2, as the PoCs are complex and ProVerif discovers  
different attack patterns  depending on the
situation.
For the WebCrypto API, one may argue that the PoC's
attack-specific parts are relatively small and only two
attack patterns are relevant. The user might therefore decide to write a custom PoC generator and
forego attack search via ProVerif.
We hold two arguments against this:
first, the a~priori knowledge about these two attack patterns exists
\emph{because} of \citeauthor{cairns2017}'s and our formal model.
Second, the concrete form of the attacks, while uniform in function,
is different because of how different key types need to be handled.
A custom PoC generator would likewise need to instantiate the abstract
attack w.r.t.\ the key types and identifiers. The PoC
generator provides a principled way of doing this. For example, adding
support for a new key type is as simples as copying and slightly
altering an existing Horn clause.

The creation of these models, including their annotations, took an
undergrad student two months. It took about the same amount of time to model the
WebCrypto API, PKCS\#11 and YubiHSM2.

    \subsection{Modeling  Guidelines}\label{modeling}

During modeling, we took note whenever we needed to adapt our models to our methodology beyond merely adding annotations.

First, 
our implementation recovers 
the clause $R$ and substitution $\sigma$ 
from a subsumed clause $R' \sqsubseteq_\sigma R$.
This method requires that all variables that appear in the conclusion
of a clause also appear in the hypothesis unless they only appear in
a name.
This means that the conclusion should be determined by
its input unless it represents a random choice.
This was no obstacle in practice, as such arbitrarily chosen variables
could just be added to the premise with an extra $\iknows$
predicate. We had to do this once, in case of the YubiHSM2 model, where
the intruder can (more or less) freely chose the attribute with which
a plain text key is stored on the device.

Second, and more fundamentally, attack recovery requires models that
are as complete as possible. That is, each attack found in a model
must represent a real-world attack. A false attack can
be translated (as our translation maps every derivation to a string)
but the PoC would fail to mount the attack.

Third, abstractions that remove variables are sometimes not applicable
because the PoC needs the information contained in this variable. A typical
example is the modeling of handles. This occurred in all three case
studies, but we refer to the running example \cref{fig:runningexample}
for illustration. From a modeling perspective, the handle, i.e., the
first parameter of the predicate \storedkey serves no purpose: the
rules for on-device computation (export and import) do not require its
knowledge. Indeed, the handle is usually guessable. Moreover, while
a key can be stored multiple times with the same attribute, this does
not give the adversary additional power (in this example; 
models where, e.g.,  the attribute can change must include the
handle~\cite{focardiFormallyVerifiedConfiguration2021,dax2019}).
Nevertheless, removing the handle makes it impossible for the PoC to
track the handle given at key generation and use the same value in
subsequent operations (i.e., those that have the \storedkey predicate
in their hypothesis).
Predicates thus need to symbolically represent the information
required by the Horn clause templates.

    \subsection{Future Work}\label{sec:futurework}

Motivated by the variety of configuration and the commercial interest
in automated PoC configuration~\cite{bortolozzo2010}, we focused on
the case of Security APIs. A natural question would be how to extend
our approach to network protocols. Fundamentally, the approach would
be similar: a derivation tree could be translated into code that
produces messages in the network. This code, however, is much more
complicated in the network case, as here the attacker cannot control
the state of the honest parties like it controls the state of the
Security API. 
Network attacks often require benign communication attempt by honest
parties, for example in
machine-in-the-middle attacks.
The PoC would need to read messages from the network, decode them and
deduce from them the internal state of the honest parties, so it sends
the attacker's
packets at the right time. This is compatible with the current
approach, but would greatly benefit from additional tooling, e.g.,
for decoding network packets into their Dolev-Yao representation.

Compared to Tookan~\cite{bortolozzo2010}, we generalized the PoC
generation, but not the reverse-engineering procedure, which polls the 
device for information to produce a Horn clause representation of the
initial state. As ProVerif (or any other verification tool) is only
called after specifying the initial state and thus defining the model, integrating the reverse-engineering
step into the verifier has no benefit. Instead, any script can produce this part of
the model. We do not see an approach for generalizing this task. 

As mentioned before, our models are not feature-complete. For
PKCS\#11, most of the literature picks a subset of the
functionality due to the sheer size of the standard.
For the WebCrypto API, we could adapt the model to analyze web
applications that hide the keys
(as discussed in \cref{sec:webcrypto} and \cite[Section~5.2]{webcryptoapi}).
For the YubiHSM model, one could adopt the domain feature, which would
be a straightforward change. 

\section{Conclusion}
\label{sec:conclusion}

We presented a method to automatically generate proof-of-concept
exploits on security APIs based on a formal model.
Although we chose to use ProVerif's typed Horn clause calculus as the
underlying basis, our approach generalizes to other logic-based
techniques, as discussed in \cref{sec:related}.
We applied this method to three very different APIs without any difficulty.
From this evaluation we learned that our template-based approach streamlines PoC generation, specifically when
the PoC has to accommodate for keys of different types. 
We also presented the first formal model of the YubiHSM2,
which features a 
very expressive security model. Unfortunately, this makes it
difficult to examine a configuration for security; a situation that
our implementation and our model mitigate in practice.

\printbibliography

@inproceedings{chevalLittleMoreConversation2018,
  title = {A {{Little More Conversation}}, a {{Little Less Action}}, a {{Lot More Satisfaction}}: {{Global States}} in {{ProVerif}}},
  shorttitle = {A {{Little More Conversation}}, a {{Little Less Action}}, a {{Lot More Satisfaction}}},
  booktitle = {2018 {{IEEE}} 31st {{Computer Security Foundations Symposium}} ({{CSF}})},
  author = {Cheval, Vincent and Cortier, Veronique and Turuani, Mathieu},
  date = {2018-07},
  pages = {344--358},
  publisher = {{IEEE}},
  location = {{Oxford}},
  doi = {10.1109/CSF.2018.00032},
  url = {https://ieeexplore.ieee.org/document/8429316/},
  urldate = {2022-09-26},
  eventtitle = {2018 {{IEEE}} 31st {{Computer Security Foundations Symposium}} ({{CSF}})},
  isbn = {978-1-5386-6680-7},
  langid = {english}
}

@inproceedings{clarkeCounterexampleGuidedAbstractionRefinement2000,
  title = {Counterexample-{{Guided Abstraction Refinement}}},
  booktitle = {Computer {{Aided Verification}}},
  author = {Clarke, Edmund and Grumberg, Orna and Jha, Somesh and Lu, Yuan and Veith, Helmut},
  editor = {Emerson, E. Allen and Sistla, Aravinda Prasad},
  date = {2000},
  series = {Lecture {{Notes}} in {{Computer Science}}},
  pages = {154--169},
  publisher = {{Springer}},
  location = {{Berlin, Heidelberg}},
  doi = {10.1007/10722167_15},
  isbn = {978-3-540-45047-4},
  langid = {english}
}

@inproceedings{clulowSecurityPKCS112003,
  title = {On the {{Security}} of {{PKCS}} \#11},
  booktitle = {Cryptographic {{Hardware}} and {{Embedded Systems}} - {{CHES}} 2003},
  author = {Clulow, Jolyon},
  editor = {Walter, Colin D. and Koç, Çetin K. and Paar, Christof},
  date = {2003},
  series = {Lecture {{Notes}} in {{Computer Science}}},
  pages = {411--425},
  publisher = {{Springer}},
  location = {{Berlin, Heidelberg}},
  doi = {10.1007/978-3-540-45238-6_32},
  isbn = {978-3-540-45238-6},
  langid = {english}
}

@inproceedings{focardiFormallyVerifiedConfiguration2021,
  title = {A {{Formally Verified Configuration}} for {{Hardware Security Modules}} in the {{Cloud}}},
  booktitle = {Proceedings of the 2021 {{ACM SIGSAC Conference}} on {{Computer}} and {{Communications Security}}},
  author = {Focardi, Riccardo and Luccio, Flaminia L.},
  date = {2021-11-13},
  series = {{{CCS}} '21},
  pages = {412--428},
  publisher = {{Association for Computing Machinery}},
  location = {{New York, NY, USA}},
  doi = {10.1145/3460120.3484785},
  url = {https://doi.org/10.1145/3460120.3484785},
  urldate = {2022-11-10},
  isbn = {978-1-4503-8454-4}
}

@article{kremerAutomatedAnalysisSecurity,
  title = {Automated Analysis of Security Protocols with Global State},
  author = {Kremer, Steve and Künnemann, Robert},
  pages = {16},
  langid = {english}
}

@online{MetasploitWebpage,
  title = {Metasploit {{Webpage}}},
  url = {https://www.metasploit.com/},
  urldate = {2022-12-09},
  langid = {english},
  organization = {{Metasploit}}
}

@article{nguyenAbstractionsSecurityProtocol,
  title = {Abstractions for {{Security Protocol Veriﬁcation}}},
  author = {Nguyen, Binh Thanh and Sprenger, Christoph and Cremers, Cas},
  pages = {115},
  langid = {english}
}

@inproceedings{rogawayProvableSecurityTreatmentKeyWrap2006,
  title = {A {{Provable-Security Treatment}} of the {{Key-Wrap Problem}}},
  booktitle = {Advances in {{Cryptology}} - {{EUROCRYPT}} 2006},
  author = {Rogaway, Phillip and Shrimpton, Thomas},
  editor = {Vaudenay, Serge},
  date = {2006},
  series = {Lecture {{Notes}} in {{Computer Science}}},
  pages = {373--390},
  publisher = {{Springer}},
  location = {{Berlin, Heidelberg}},
  doi = {10.1007/11761679_23},
  isbn = {978-3-540-34547-3},
  langid = {english}
}

@online{StartupsResultingProject,
  title = {Startups Resulting from the Project Teams of the {{Inria Saclay}} - {{Île-de-France}} Research Centre | {{Inria}}},
  url = {https://www.inria.fr/en/startups-resulting-project-teams-inria-saclay-ile-de-france-research-centre},
  urldate = {2022-11-10}
}

@misc{yubicoabYubiHSMYubiHSMFIPS,
  title = {{{YubiHSM}} 2 and {{YubiHSM}} 2 {{FIPS}}: {{Product}} Brief},
  author = {{Yubico AB}},
  url = {https://resources.yubico.com/53ZDUYE6/at/q4bsft-z2wi8-fo7aqg/YubiHSM2_Product_Brief.pdf?format=pdf}
}

@online{secapifaq,
  title={{Analysis of Security APIs FAQ}},
  author={Graham Steel},
  url={http://www.lsv.fr/~steel/security_APIs_FAQ.html},
  note = {[Accessed: 17.09.2019]}
}

@online{pkcs11,
  title={{PKCS \#11 Cryptographic Token Interface Base Specification Version 2.40 Plus Errata 01}},
  url={http://docs.oasis-open.org/pkcs11/pkcs11-base/v2.40/pkcs11-base-v2.40.html},
  note = {[Accessed: 11.09.2019]}
}

@online{webcryptoapi,
  title={{Web Cryptography API}},
  url={https://www.w3.org/TR/WebCryptoAPI/},
  note = {[Accessed: 11.09.2019]}
}

@online{webcryptosupport,
  title={{Web Crypto API}},
  url={https://developer.mozilla.org/en-US/docs/Web/API/Web_Crypto_API},
  note = {[Accessed: 02.08.2019]}
}

@online{mscng,
  title={{Cryptography API: Next Generation}},
  url={https://docs.microsoft.com/de-de/windows/win32/seccng/cng-portal},
  note = {[Accessed: 11.09.2019]}
}

@online{yubihsm,
  title={{YubiHSM 2}},
  url={https://www.yubico.com/products/yubihsm/},
  note = {[Accessed: 11.09.2019]}
}

@online{yubihsmdocu,
  title={{YubiHSM 2}},
  url={https://developers.yubico.com/YubiHSM2/},
  note = {[Accessed: 17.09.2019]}
}

@inproceedings{fröschle2011,
  title={{Reasoning with Past to Prove PKCS\#11 Keys Secure}},
  author={Fröschle, Sibylle and Sommer, Nils},
  booktitle={Formal Aspects of Security and Trust},
  pages={96--110},
  year={2011},
  organization={Springer}
}

@inproceedings{adao2013,
 title = {{Type-Based Analysis of Generic Key Management APIs}},
 author = {Adao, Pedro and Focardi, Riccardo and Luccio, Flaminia L.},
 booktitle = {Proceedings of the 2013 IEEE 26th Computer Security Foundations Symposium},
 pages = {97--111},
 year = {2013},
 publisher = {IEEE Computer Society}
}

@inproceedings{tamarin,
  title={{The TAMARIN Prover for the Symbolic Analysis of Security Protocols}},
  author={Meier, Simon and Schmidt, Benedikt and Cremers, Cas and Basin, David},
  booktitle={Computer Aided Verification},
  pages={696--701},
  year={2013},
  publisher={Springer}
}

@inproceedings{dax2019,
  title={{How to Wrap it up - A Formally Verified Proposal for the use of Authenticated Wrapping in PKCS\#11}},
  author={Dax, Alexander and Künnemann, Robert and Tangermann, Sven and Backes, Michael},
  booktitle={2019 IEEE 32nd Computer Security Foundations Symposium (CSF)},
  year={2019},
  url={https://publications.cispa.saarland/2895/},
  note={[Accessed: 07.09.2019]}
}

@inproceedings{künnemann2015,
  title={{Automated Backward Analysis of PKCS\#11 v2.20}},
  author={Künnemann, Robert},
  booktitle={Principles of Security and Trust},
  pages={219--238},
  year={2015},
  publisher={Springer}
}

@inproceedings{gonzalezburgueno2018,
  title={{Formal verification of the YubiKey and YubiHSM APIs in Maude-NPA}},
  author={González-burgue\~no, Antonio and Aparicio-Sánchez, Damián and Escobar, Santiago and Meadows,Catherine and Meseguer, José},
  booktitle={LPAR-22. 22nd International Conference on Logic for Programming, Artificial Intelligence and Reasoning},
  volume={57},
  pages={400--417},
  year={2018},
  publisher={EasyChair}
}

@incollection{BlanchetBook09,
  author={Bruno Blanchet},
  title={{Using Horn Clauses for Analyzing Security Protocols}},
  booktitle={Formal Models and Techniques for Analyzing Security Protocols},
  publisher={IOS Press},
  pages={86--111},
  volume={5},
  series={Cryptology and Information Security Series},
  month={MAR},
  year={2011},
  editor={V{\'e}ronique Cortier and Steve Kremer},
  ISBN={978-1-60750-713-0}
}

@inproceedings{bortolozzo2010,
  title={Attacking and Fixing PKCS\#11 Security Tokens},
  author={Bortolozzo, Matteo and Centenaro, Matteo and Focardi, Riccardo and Steel, Graham},
  booktitle={Proceedings of the 17th ACM Conference on Computer and Communications Security},
  pages={260--269},
  year={2010},
  publisher={ACM}
}

@inproceedings{künnemannsteel2013,
  title={{YubiSecure? Formal Security Analysis Results for the Yubikey and YubiHSM}},
  author={Künnemann, Robert and Steel, Graham},
  booktitle={Security and Trust Management},
  pages={257--272},
  year={2013},
  publisher={Springer}
}

@inproceedings{cairns2017,
  title={{Security Analysis of the W3C Web Cryptography API}},
  author={Cairns, Kelsey and Halpin, Harry and Steel, Graham},
  booktitle={{Proceedings of Security Standardisation Research (SSR)}},
  series={Lecture Notes in Computer Science (LNCS)},
  volume={10074},
  pages={112 - 140},
  year={2017},
  month=Dec,
  publisher={{Springer}}
}

@online{proverif,
  title={{ProVerif: Cryptographic protocol verifier in the formal model}},
  url={https://prosecco.gforge.inria.fr/personal/bblanche/proverif/},
  note={[Accessed: 11.09.2019]}
}

@online{pythonpkcs11,
  title={{Python PKCS\#11 - High Level Wrapper API}},
  url={https://python-pkcs11.readthedocs.io/en/latest/index.html},
  note={[Accessed: 12.09.2019]}
}

@online{opencryptoki,
  title={{openCryptoki}},
  url={https://github.com/opencryptoki/opencryptoki},
  note={[Accessedd: 12.09.2019]}
}

@online{pythonyubi,
  title={python-yubihsm},
  url={https://developers.yubico.com/python-yubihsm/},
  note={[Accessed: 12.09.2019]}
}

@inproceedings{delaune2008,
  title={{Formal Analysis of PKCS\#11}},
  author={Delaune, Stéphanie and Kremer, Steve and Steel, Graham},
  booktitle={Proceedings of the 2008 21st IEEE Computer Security Foundations Symposium},
  pages={331--344},
  year={2008},
  publisher={IEEE Computer Society}
}

@article{bond01api,
  Author =	 {M. Bond and R. Anderson},
  Journal =	 {IEEE Computer Magazine},
  Month =	 oct,
  Pages =	 {67-75},
  Title =	 {{API} level attacks on embedded systems},
  Year =	 2001
}

@article{longley92automatic,
  Author =	 {Dennis Longley and Simon Rigby},
  Journal =	 {Computers and Security},
  Month =	 mar,
  Number =	 1,
  Pages =	 {75--89},
  Publisher =	 {Elsevier},
  Title =	 {An Automatic Search for Security Flaws in Key
                  Management Schemes},
  Volume =	 11,
  Year =	 1992
}

@article{DKS-jcs09,
  Author =	 {Delaune, St{\'e}phanie and Kremer, Steve and Steel,
                  Graham},
  Doi =		 {10.3233/JCS-2009-0394},
  Journal =	 {Journal of Computer Security},
  Month =	 nov,
  Number =	 6,
  Pages =	 {1211-1245},
  Publisher =	 {{IOS}},
  Title =	 {Formal Analysis of {PKCS\#11} and Proprietary
                  Extensions},
  Url =
                  {http://www.lsv.ens-cachan.fr/Publis/PAPERS/PDF/DKS-jcs09.pdf},
  Volume =	 18,
  Year =	 2010
}

@inproceedings{KKS-esorics13,
	Author = {Kremer, Steve and K{\"u}nnemann, Robert and Steel, Graham},
	Booktitle = {ESORICS 2013},
	OPTEditor = {Crampton, Jason and Jajodia, Sushil and Mayes, Keith},
	OPTPages = {327--344},
        series    = {LNCS},
        volume    = {8134},
	Publisher = {Springer},
	Title = {Universally Composable Key-Management},
	Year = {2013}}

@inproceedings{Kremer2011Security-for-Ke,
	Author = {Kremer, Steve and Steel, Graham and Warinschi, Bogdan},
	Booktitle = {CSF 2011},
	Pages = {66--82},
	publisher = {IEEE Computer Society},
	Title = {Security for Key Management Interfaces},
	Year = 2011
}

@inproceedings{Cortier:2012:RLL:2382196.2382293,
 author = {Cortier, V{\'e}ronique and Steel, Graham and Wiedling, Cyrille},
 title = {Revoke and let live: a secure key revocation {API} for cryptographic devices},
 booktitle = {CCS 2012},
 year = {2012},
 OPTpages = {918--928},
 publisher = {ACM},
}

@inproceedings{cortier07TACAS,
	Author = {V. Cortier and G. Keighren and G. Steel},
	Booktitle = {TACAS 2007},
	OPTPages = {538--552},
        series    = {LNCS},
          volume    = {4424},
            publisher = {Springer},
	Title = {Automatic Analysis of the Security of {XOR}-based Key Management Schemes},
	Year = 2007}

\appendix
\subsection{Statement on Data Availability}

We will publish our models and make our tool's source code available
under the conditions of the GPL.

\subsection{YubiHSM: Reverse engineering
notes}\label{sec:yubihsm-reverseengineering}

The YubiHSM is a hardware security module that was mainly designed for protecting cryptographic keys. These keys can be used to perform cryptographic operations on the device, which supports both symmetric and asymmetric cryptography. Further, since version 2 it supports key wrapping, which allows to internally encrypt (``wrap'') keys on the device with other keys (called Wrap Keys). The resulting encryption is then given to the user, and can be used to import the wrapped key (possibly on another device) using the same key that was used for encryption.

There are five types of keys: Authentication Keys are used to establish sessions between the device and the users. Wrap Keys are used for key wrapping. OTP-AEAD keys are used for authenticated encryption, and play an important role in Yubico's OTP protocol, which is also used by the YubiKey. Asymmetric Keys are private keys that can be used for asymmetric decryption and signatures. Finally, HMAC Keys can (unsurprisingly) be used to perform HMAC operations.

In addition to the type, each key has a two byte ID as well as other attributes. The combination of ID and key type are used to identify keys, so there can be two keys with the same ID, but not with the same ID AND the same type. Moreover, each key has a label, a set of domains, and a set of capabilities. A label is a string of 40 bytes that can be used to store information about a key. Domains are used to separate keys from one another and can roughly be seen as partitions. There are 16 domains, and each key is assigned to at least one or more of them. Keys can only interact with each other when they have at least one domain in common. Further, a session also has a set of domains which is inherited by the Authentication Key used to establish it, and also needs to share at least one domain with a key so that it can use it. Capabilities are used to define what keys can be used for, so e.g. a Wrap Key can only be used for wrapping if it has the capability ``export-wrapped''. Sessions also have capabilities, which are again inherited by the Authentication Keys, and define what operations can be performed in a session. Additionally, Wrap Keys and Authentication Keys also have a set of delegated capabilities. For Wrap Keys, those define the capabilities that keys are allowed to have when they are imported or exported with a specific Wrap Key. For Authentication Keys, they are once again inherited by the session and define the set of capabilities that can be assigned to other keys created in that session.

To interact with the device, the YubiHSM offers a vendor-specific API. In our analysis, we only focused on a small subset of commands since those are the only ones relevant for our approach.

We focused only on commands that are relevant to key management, while also ignoring some commands that are related to key management but do not give us any advantage when trying to extract keys that are already on the device. For example, ``Put Hmac Key'' is a key management command, but since HMAC Keys are in no way related to the wrapping functionality, being able to create those keys does not help us in extracting other keys. 
One thing about the HSM that is also important for our approach is the internal format of key wrappings. That is, the format in which information about a key is stored in the ciphertext that results from a wrapping operation. This format is not mentioned in the official documentation, but we determined it by reverse engineering. It looks as follows:
\begin{itemize}
\item 1 Byte Algorithm of Wrap Key
\item 8 Bytes Capabilities
\item 2 Bytes ID
\item 2 Bytes Size
\item 2 Bytes Domains
\item 1 Byte Type
\item 1 Byte Algorithm
\item 1 Byte Sequence
\item 1 Byte Origin
\item 40 Bytes Label
\item 8 Bytes Delegated Capabilities (only for Authentication Keys and Wrap Keys)
\item 4 Bytes nonce ID (only for OTP AEAD Keys)
\end{itemize}
After all that comes the actual cryptographic key. The first byte is especially important, as it is also used to ensure that a key wrapping cannot be decrypted with the ``Unwrap Data'' command. The reason for that is that the ``Wrap Data'' command always sets this byte to 0, so the ``Unwrap Data'' command can check by this byte if a wrapping originates from a call to ``Wrap Data''. If not, the wrapping is rejected. Another thing we learned from the analysis of this format is how the HSM computes the size of a key. After working with the device for a while, we noticed that when using ``Get Object Info'' to get information about a key, the size value does not alway fit to the length that the key should actually have. However, by closer examining the wrapping format, we were able to determine the reason for that: When the size of a key is computed, the device always looks at the internal representation of the key (which seeems to have the same format as the wrappings) and counts all bytes that come after the label, expecting that they all belong to the key. However, this is not true for Authentication Keys, Wrap Keys and OTP AEAD Keys, as they all have specific fields that come after the label but before the actual key. Therefore, the computed size values for those keys are wrong. Moreover, HMAC Keys are another special case. For those we not only noted that the size value is wrong, but also that the wrapped keys were different from what we specified when using ``Put Hmac Key''. The reason for that is that Hmac keys are not stored in plain, but instead as the result of the first step of an HMAC computation, which is always the same and only depends on the key. This encoding led to the size value always being twice as much as the specified key length.

Furthermore, we also examined the HSM using fuzzing. For that we built two fuzzers, one that was designed to create valid wrappings and try to import them to see if some of them cause errors, and one designed to create invalid wrappings, to see if some of them could be imported despite having errors. While this fuzzing did not show as anything unusual or dangerous, we noticed some oddities that are worth mentioning: First, we noticed that the Origin byte allows more values than are actually needed. This byte is meant to determine where an object on the device originally came from. 0x01 means the object has been generated on the device, 0x02 means the object has been imported using the put command, and 0x10 means the object has been imported via wrapping. These can also be combined, such that there are four valid values for the origin: 0x01, 0x02, 0x11 and 0x12. However, when crafting a wrapped object and importing it, we noticed that any other value is also accepted. This byte will then be assigned to the resulting object exactly as specified, with the exception of the bit that indicates the object was imported, which is always set to 1 by the device.
Another thing that came to our attention was something that we actually found out while writing the fuzzer. That is, we noticed that there is a maximum length for messages sent to the device, which is mentioned briefly in the documentation but it is not really highlighted. We found out that this limit leads to a corner case where it is not possible to wrap an object even though it is marked as exportable. This can happen when usig the command ``Put Opaque'', which allows to store arbitrary bytes on the device as an opaque object. When the amount of bytes sent with this command is barely small enough that the message as a whole is not too long, the object will be correctly stored on the device, but wrapping will not work because the wrapped object becomes longer than the allowed message size. The object is still accessible via the command ``Get Opaque'', so this does not have a huge impact, but it is still slightly odd.


\end{document}  